\documentclass[12pt, twoside]{article}
\usepackage[utf8]{inputenc}
\usepackage{setspace}
\usepackage{fancyhdr}
\usepackage[margin=1in, headheight=15pt]{geometry}
\usepackage[dvipsnames]{xcolor}
\usepackage{graphics}
\usepackage{tikz}
\usetikzlibrary{calc}
\usetikzlibrary{decorations.pathreplacing}
\usetikzlibrary{arrows.meta}
\tikzset{every picture/.style={/utils/exec={\sffamily}}}
\usepackage{amsmath,amsfonts,amsthm}
\usepackage{mathrsfs}
\usepackage{cancel}
\usepackage[normalem]{ulem}
\usepackage{siunitx}
\usepackage{float}
\usepackage{enumitem}
\usepackage{colortbl}
\usepackage{tabularx}
\usepackage{todonotes}
\newcolumntype{L}[2]{>{\hsize=#1\hsize\columncolor{#2}\raggedright\arraybackslash}X}%
\newcolumntype{R}[2]{>{\hsize=#1\hsize\columncolor{#2}\raggedleft\arraybackslash}X}%
\newcolumntype{C}[2]{>{\hsize=#1\hsize\columncolor{#2}\centering\arraybackslash}X}%
\usepackage{wrapfig}
\usepackage{transparent}
\usepackage{booktabs}
\usepackage{subcaption}
\usepackage[font=footnotesize,labelfont=bf]{caption}
\usepackage{titlesec}
\usepackage[nottoc,numbib]{tocbibind}
\usepackage{tabularx}
\usepackage[hang,flushmargin]{footmisc} 

\usepackage{lineno}

\usepackage{natbib}
\setcitestyle{aysep={}}

\usepackage{hyperref}

\hypersetup{colorlinks=true, citecolor=gray, linkcolor=gray, urlcolor=gray}

\makeatletter
\newcommand{\myfnsymbol}[1]{%
  \expandafter\@myfnsymbol\csname c@#1\endcsname
}
\newcommand{\@myfnsymbol}[1]{%
  \ifcase #1
  \or 1
  \or 2
  \or 3
  \or 4
  \or \TextOrMath{\textasteriskcentered}{*}
  \or \TextOrMath{\textdagger}{\dagger}
  \or \TextOrMath{\textdaggerdbl}{\ddagger}
  \fi
}
\newcommand{\affiliationA}{\@myfnsymbol{1}}
\newcommand{\affiliationB}{\@myfnsymbol{2}}
\newcommand{\affiliationC}{\@myfnsymbol{3}}
\newcommand{\affiliationD}{\@myfnsymbol{4}}
\newcommand{\titlenote}{\@myfnsymbol{5}}
\newcommand{\equalcontrib}{\@myfnsymbol{6}}
\newcommand{\corresp}{\@myfnsymbol{7}}
\makeatother

\newcommand{\mytitle}{Incivility in Public Health Policy Discussions Spills Over to Public Engagement with Climate Issues}
\newcommand{\shorttitle}{Incivility Spillover to Public Engagement with Climate Issues}

\author{
Hasti Narimanzadeh\textsuperscript{\affiliationA,\equalcontrib}
\and Arash Badie-Modiri\textsuperscript{\affiliationA,\equalcontrib}
\and Iuliia Smirnova\textsuperscript{\affiliationB}
\and Ted Hsuan Yun Chen\textsuperscript{\affiliationC,\corresp}
}

\date{}

\title{\mytitle}
\begin{document}
\pagenumbering{roman}
\singlespacing

\renewcommand{\thefootnote}{\myfnsymbol{footnote}}
\maketitle\thispagestyle{empty}
\footnotetext[1]{Department of Computer Science, Aalto University, Finland.}%
\footnotetext[2]{Faculty of Social Sciences, University of Helsinki, Finland.}%
\footnotetext[3]{Department of Environmental Science and Policy, George Mason University, USA.}%
\footnotetext[6]{Equal contribution ordered by reverse seniority.}%
\footnotetext[7]{Corresponding author. Email: ted.hsuanyun.chen@gmail.com. Address: 3036 David King Hall, 4400 University Drive MSN: 5F2. Fairfax, Virginia. 22030. USA.}%

\begin{abstract}
    \noindent Affective polarization and political sorting drive public antagonism around climate change and other issues at the science-policy nexus. We study cross-domain spillover of incivility in public engagements with climate change and public health on Twitter and Reddit using the COVID-19 period as a case study. We find strong evidence of the signatures of affective polarization surrounding COVID-19 spilling into the climate change domain. Across different social media systems, COVID-19 content is associated with incivility in climate discussions. These patterns of increased antagonism were responsive to pandemic events that made the link between science and public policy more salient. The observed spillover activated along pre-pandemic political cleavages, specifically anti-internationalist populist beliefs, that linked climate policy opposition to vaccine hesitancy. Our findings show how affective polarization in public engagement with science becomes entrenched across policy domains, which has implications for how the public engages with and communicates about issues such as climate change and public health.
    \vspace{0.5cm}
    
    \vspace{0.5cm}

\end{abstract}


\clearpage

\onehalfspacing
\pagenumbering{arabic}
\setcounter{page}{1}

\section{Introduction}
There is a virtual consensus in the scientific community that climate change is caused by human activity~\citep{myers2021consensus,cook2013quantifying}, and an overwhelming majority of scientists agree that fundamental changes are needed to address the climate crisis~\citep{dablander2024climate}. Still, a non-trivial minority of the public remains skeptical of climate science and is opposed to stronger climate policies~\citep{leiserowitz2023climate}. Notably, they are vocal about their skepticism and opposition on social media~\citep{falkenberg2022growing,rojas2024hierarchical}, often in a highly uncivil manner, which begets further antagonistic intergroup conflict~\citep{xia2021spread,xia2020exploring}. A similar phenomenon exists in the public health domain, which became widespread during the COVID-19 pandemic when highly antagonistic anti-vaccine and anti-science sentiments were rampant on social media~\citep{hsieh2025psychological,chen2023anti}. Early evidence suggests these trends are continuing into the post-pandemic period~\citep{poddar2024covid}.

Discussions over empirical findings and policy positions are the cornerstones of the scientific endeavor and the democratic process, but research from the science and political communication fields shows that antagonistic and uncivil interactions are detrimental to meaningful engagement \citep{van2024opposing}, especially when positions are taken based on pre-existing societal cleavages instead of the merits of the issue itself \citep{druckman2025partisan}.
Incivility, which ranges from incensed discussions to outright rude critiques and name-calling~\citep{anderson2014nasty}, intensifies intergroup polarization~\citep{brundidge2024clinching, chen2024disagreement}. It is associated with the dehumanization of outgroup members~\citep{moore2020exaggerated}, reducing individuals' willingness to engage with each other~\citep{ng2020toward}, and lower trust in institutions~\citep{mutz2005new}. Specifically in terms of public trust of science, incivility has been found to delegitimize science and lower public trust toward scientists \citep{chinn2022can}.


%

In this study, we explore the public's uncivil and antagonistic engagement with climate science and public health discussions, two critical issues at the science-policy nexus. Different from prior research on incivility of discussion in climate change \citep{xia2021spread} and public health \citep{hsieh2025psychological}, which primarily looked at the issues by themselves, we focus on how outgroup antagonism spills from one issue domain to another.


We conceptualize spillover of antagonism between societal issues as a manifestation of affective polarization resulting from individuals sorting into entrenched political groups~\citep{mason2015disrespectfully}. There is a growing tendency for issues to become linked through their association with partisan positions, whereby individuals' attitudes toward different issues become predictive of each other~\citep{kozlowski2021issue}. As more issues become aligned in this manner, entrenched group-based cleavages become salient cues for belief formation and political behavior \citep{baldassarri2008partisans}, including how one understands new policy issues and engages with their outgroup \citep{taber2006motivated,druckman2019evidence}. When a highly aligned system is also affectively polarized, antagonistic communication cues and outgroup mistrust in one domain should easily travel to another. Following this mechanism, linkages between issues that share overlapping sources of political cleavage can also become a pathway for the spreading of political extremism \citep{miller2020covid,enders2021conspiracy}. Politically-linked conspiracy beliefs, for example, is a particularly salient concern in the public health domain \citep{erokhin2022covid}.

To study public health spilling over to climate change, we look at the COVID-19 period, when the severity of the pandemic increased the visibility and importance of communication between the general public and the scientific community on public policy matters. This period saw high levels of politicization and polarization toward COVID-19~\citep{green2020elusive, hart2020politicization}, which likely contributed to the diverging trust in science, including scientists and states' scientific institutions, among different segments of the population~\citep{levesque2025contesting,radrizzani2023both, kerr2021political}. The impact of the COVID-19 shock therefore provides us an opportune case study to examine important patterns in how the public engages with science across different issues, as novel events from the pandemic can reveal evidence of spillover from public health to climate change discussions. 

We explore the following questions about how cross-domain spillover leads to greater affective polarization in the public's engagement with science. How do large scale shocks to the role of science in society change how the public engages with scientific issues across different domains? Do negative sentiments toward COVID-19 science and policy spill over to public engagement with climate science? What are the conceptual pathways through which these spillovers travel? 
With the constant spotlight on the role of science in policymaking during the pandemic, we expect public tendencies surrounding COVID-19 to spill into how the public engages with climate science. Specifically, given the salience of polarized elite political cues surrounding COVID-19~\citep{green2020elusive}, we expect increased incivility in climate discussions that also involve COVID-19 topics. 

We approach our research questions by studying antagonistic public engagement with COVID-19 and climate change-related content on social media platforms characterized by a high volume of cross-group science and policy debates.
We focus on Twitter, which was an important space for public contestations over social and political issues~\citep{theocharis2015conceptualization}, and was recognized within the scientific community as the primary social media platform for science communication~\citep{hammer2021social}, making it an important case for questions about the science-public interface~\citep{walter2019scientific}. We also examine Reddit discussions to supplement our findings. We show that the presence of COVID-19 content is strongly predictive of climate change conversations becoming uncivil, that these patterns respond to landmark events in the COVID-19 domain, and that incivility spillover activates along existing sources of political polarization.

Our study makes a number of contributions to the literature on how the public engages with issues at the science-policy nexus. First, specific to how COVID-19 impacted the public's engagement with the climate issue, prior studies have primarily focused on shifts in the public's attention levels, consistently identifying an attention shift away from the climate issue that has been attributed to cognitive limits in attention or concern \citep{smirnov2022covid,sisco2023examining,stoddart2023competing,rauchfleisch2023covid,repke2024global}. In terms of policy attitude shifts, one study uncovered the potential for synergies between COVID-19 and climate change to improve policy support for both issues \citep{bergquist2023politics}. However, studies have yet to consider the possibility for spillover of polarization between the issues through harmful types of ``synergies'' that arise from existing political cleavages.
Second, in establishing these spillover patterns, we not only uncover the potential for harmful linkages, but also speak to the political attitudes research on how policy positions become constrained within increasingly polarized belief systems \citep{kozlowski2021issue,chen2021polarization} and research on how belief in different conspiracy theories reinforce one another \citep{enders2021conspiracy}.
Finally, whereas most studies on science communication in social media space are limited to single platforms, we look at both Twitter and Reddit, and find comparable results across the different systems. 


\section{Methods}

In this section, we introduce our empirical approach, and our data collection and measures. Technical details of our methodology are presented in the Supplementary Materials.
We conducted all analyses in \texttt{R} \citep{r2024}. All code and data required for reproducing our results will be made publicly available at [redacted Zenodo repository].

\subsection{Empirical Approach}


To assess whether there is spillover of affective polarization between the two science policy domains of COVID-19 and climate change, we examined how the incivility of climate posts from a 134-week period (February~2019--August~2021) varied by the presence of COVID-19. Specifically, we analyzed whether climate post incivility depend on whether they coincided with the pandemic period (March 2020 onward) and whether they contained references to COVID-19. 
We also focused particularly on references to Anthony Fauci, who, as then-Chief Medical Advisor to the President of the U.S., polarized COVID-19 discussions by being emblematic of how scientific institutions impact public policy~\citep{chen2023anti}. 

We examined temporal trends in incivility across our entire data period, focusing on how public engagements with climate change changed with the pandemic's onset and with notable pandemic period events. This examination additionally provides a data-driven method for us to focus our statistical analysis of topic co-occurrence and antagonistic engagement to after affective polarization surrounding COVID-19 stabilized following an initial escalation period marked by the politicization of the pandemic \citep{green2020elusive}.
For our statistical models of climate post incivility, we looked at co-occurrence and onset using linear probability models and survival models, where the main predictor of interest is COVID-19 content. 

By examining the intersection of these two public policy issues, we are able to uncover shifts in polarization that arise when global crises intertwine with wider debates at the science-policy nexus. Due to the variety of models we fit, we describe each model in their corresponding results section. We report the sample sizes for our models, which range between 39,000--37 million, in Supplementary Materials S5.

\subsection{Data Collection}
We used five data sets of social media posts, summarized in Table~\ref{tab:datasets}.
First, we collected 38.4 million tweets that contain keywords related to climate change. These keywords are summarized in Supplementary Materials~S1. For 287,000 of these tweets, we further collected the entirety of their ensuing conversations. We also collected the same data for tweets directly referencing climate science publications through hyperlinks, resulting in 259,000 tweets and 38,000 conversations. 
We collected all tweets using the full archive search endpoint from Twitter’s v2 API suite during October 2021--June 2022. To ensure consistency in our measures, most of which rely on natural language processing, we only used English tweets.
Additional Twitter data collection details are in Supplementary Materials~S1.
Finally, we obtained all 2.1 million comments on Reddit that mention the phrase ``climate change'' during this period, which was published by \citet{socialgrep2022reddit}. Details of our Reddit data filtering are in Supplementary Materials~S2.


\begin{table}[t]
    \centering\footnotesize\renewcommand{\arraystretch}{1.2}
\begin{tabularx}{\textwidth}{l X r r} \hline\hline
    Data Set  & Description  & Size & Users \\ \hline
     General climate tweets & Tweets published in the data collection window containing at least one climate change related keyword or hashtag & 38.4M & 5.5M \\
     Climate science tweets & Tweets containing a hyperlink to a scientific publication containing climate related keywords in its title. Data set curated using Altmetric's API. & 259K & 79K \\
     Reddit climate  & Comments containing the phrase ``climate change'', labeled by Perspective to contain English. Data set provided by \citet{socialgrep2022reddit}.  & 2.1M &  \\
     General climate conversations & Full reply tree of 287,000 conversations (stratified random sample of approximately 2100 per week) from the \emph{General climate tweets} data set that have at least one reply and two unique participants. & 1.9M & \\
     Climate science conversations & Full reply tree of all 38,000 conversations from the \emph{Climate science tweets} data set that have at least one reply and two unique participants. & 181K & \\
     \hline
    \end{tabularx}
            \caption{\textbf{Summary of all data sets used.}}\label{tab:datasets}
\end{table}

\subsection{Measurement}
Our empirical approach requires that we identify the occurrence of different public policy content and antagonistic language use. Here, we summarize how we obtained these measures.

\subsubsection{Incivility} 
To measure incivility, we labeled our climate posts with a probability of exhibiting toxicity using the Perspective API provided by Google~\citep{wulczyn2017ex,jigsaw2017perspective}. It defines toxicity to be ``a rude, disrespectful, or unreasonable comment that is likely to make you leave a discussion''~\citep{mitchell2019model}. This operationalization maps directly onto the disengaging effects of incivility that we want to capture.
The Perspective toxicity detection model is based on the Unified Toxic Content Classification multilingual architecture~\citep{lees2022new}, and is trained on data from online forums such as comments from Wikipedia talk pages~\citep{wulczyn2017ex} and New York Times, labeled by three to ten crowdsourced human annotators. 

\subsubsection{Topic Classification}
Based on the presence of specific keywords, we assigned all climate posts binary labels for the following categories: 
1) COVID-19, 2) Anthony Fauci, and 3) international organizations and elites. The keyword lists for these policy areas are presented in Supplementary Materials~S3.
We also labeled all tweets for whether they contain climate obstructionist claims using the Augmented CARDS (Computer Assisted Recognition of Denial \& Skepticism) model from~\citet{rojas2024hierarchical}, which we describe in Supplementary Materials S3.

\section{Results and Discussion}
\subsection{Incivility Spills Over between COVID-19 and Climate Change}\label{sec:toxicity-polarization}

\begin{table}[!t]
    \centering\footnotesize\renewcommand{\arraystretch}{1.5}

    \begin{tabularx}{\textwidth}{X | X} \hline\hline
 Opposing Climate and COVID-19 Policies & Supporting Climate and COVID-19 Policies  \\ \hline

Bullshit. You and your cronies are dreaming of injecting climate crisis rhetoric into a bill for COVID-19. You are as despicable as that damn commie you support. Won't be long now till congress is back red. Thank you for being a freaking idiot.
 & No surprise, all the assholes demanding \#COVID19 projection models are the same ones denying the science behind the models on global climate change. Bunch of fuckers. \\ \hline

When you find out climate change scientists are all like Dr.~Fauci. Phoney flakes who don’t know shit from shinola. & I swear to god if you dumb ass Republicans start brushing off coronavirus like global warming we're truly fucked. \\ \hline

Are people really that stupid? Do they really believe this crap? Climate change is a bunch of bullshit so is Dr. Fauci. It's just pathetic. And this figures just every now and then they're gonna make people lock down and do the mask bullshit again. Weak minded fucking sheep.
&

Too many people are so stupid to believe the idiotic conspiracies that ``global warming is a myth'' and ``COVID-19 doesn't exist'' ... Like how fucking stupid y'all got to be when people are still dying from it, people still catch it.
\\

\hline
    \end{tabularx}
            \caption{\textbf{Examples of uncivil climate tweets that reference topics related to COVID-19.} Toxicity probabilities of these tweets exceed 0.65. Tweet texts have been edited for typos and formatting.}\label{tab:toxic-tweets}
\end{table}

From examples of toxic tweets in Table~\ref{tab:toxic-tweets}, we see that both COVID-19 content was used in polarized, vitriolic debates on climate science. It is also evident that while prior work identified incivility to be more prevalent among individuals who oppose climate policies \citep{xia2021spread}, it is not a one-sided phenomenon. Instead, supporters and opponents both use highly toxic language toward their outgroup while invoking concerns about COVID-19. 
%

\begin{figure}[!t]
    \centering
    \includegraphics[width=1\textwidth]{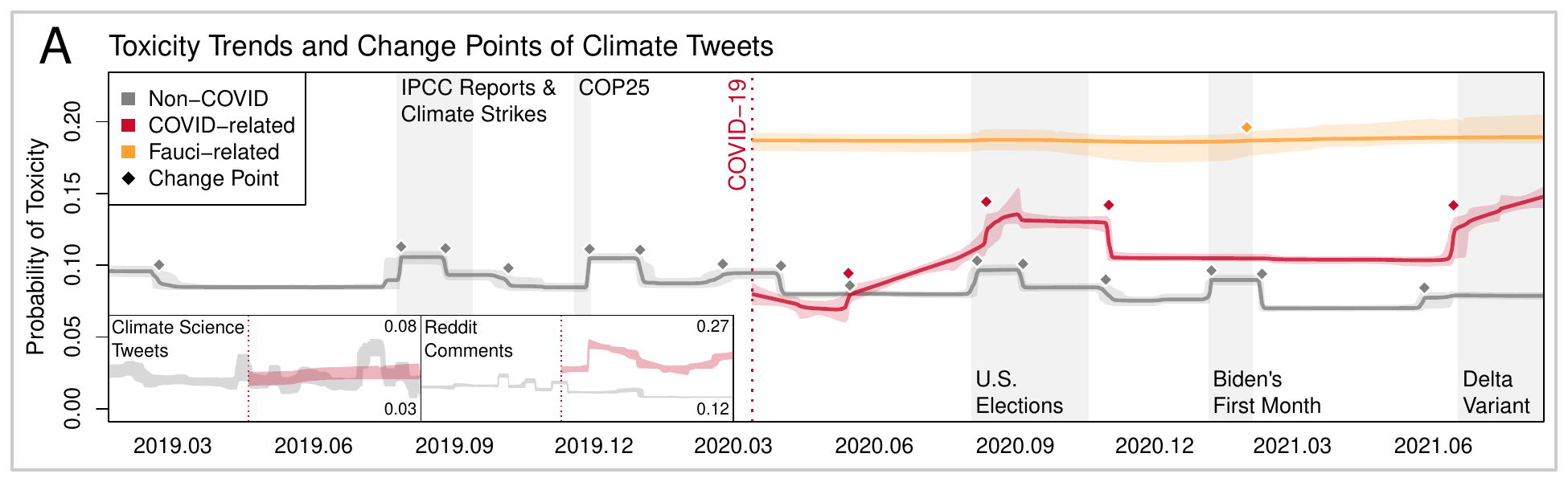}
    \includegraphics[width=1\textwidth]{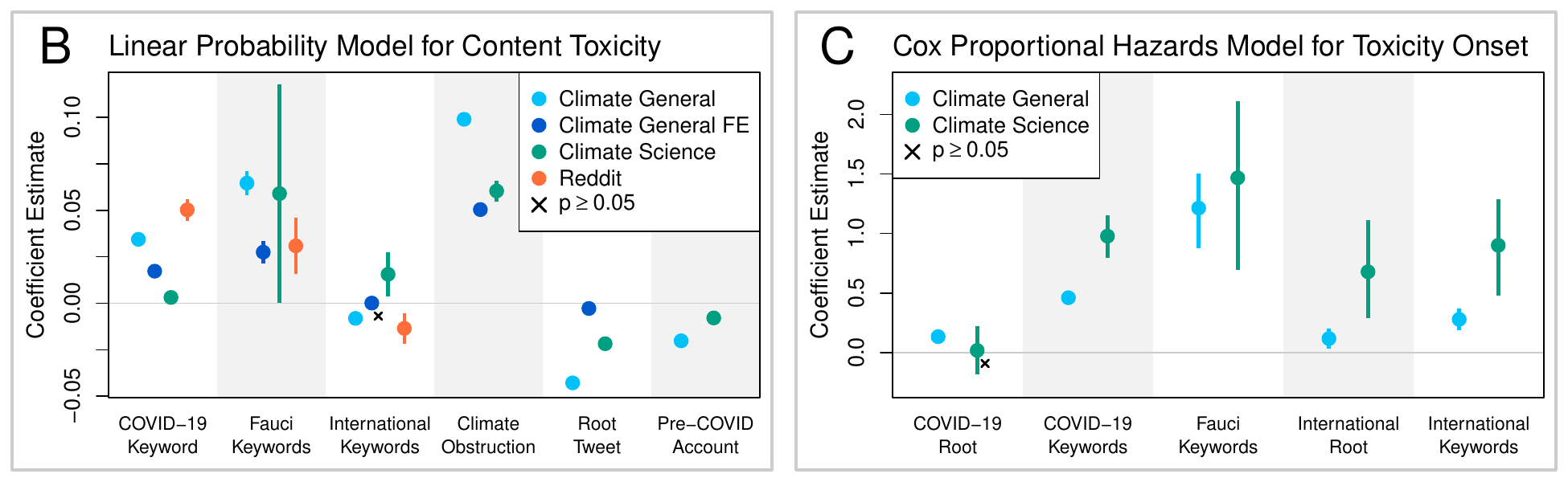}
    \caption{\textbf{Presence of COVID-19 content and incivility of climate posts.} \textbf{A.} Temporal trends in daily climate tweet toxicity by difference in content. Inset shows the temporal trends for climate science tweets and Reddit comments. \textbf{B.} Coefficient estimates from linear probability model of toxicity in climate posts across three social media systems. \textbf{C.} Coefficient estimates from Cox proportional hazards models of toxicity onset in climate conversations on Twitter.}
    \label{fig:toxicity}
\end{figure}

\subsubsection{Temporal Analysis of Incivility Dynamics}
We begin by looking at broader temporal trends in antagonistic engagement. We used the Bayesian Estimator of Abrupt change, Seasonal change, and Trend (BEAST) model \citep{zhao2019detecting}, estimating toxicity trends for climate posts with and without COVID-19 and Fauci content in all three data sets -- general climate Twitter, climate science Twitter, and climate Reddit.

BEAST is a trend and change point estimation model designed for noisy and cyclical data~\citep{zhao2019detecting}, such as social media patterns~\citep{reuning2022facebook}. It does so using a time series decomposition algorithm designed to analyze nonlinear temporal dynamics across multiple timescales while taking into account seasonal behavior (e.g.,~weekly trends). For our purposes, it allows us to observe the overall temporal patterns of antagonism spillover prior to and during the COVID-19 period, including detecting the most likely timing of affective polarization having stabilized.\footnote{BEAST is implemented in the \texttt{Rbeast} package \citep{zhao2019detecting}. A technical summary of BEAST is in Supplementary Materials S4.}

Our results are shown in Figure~\ref{fig:toxicity}.A. Broadly, we find a generally strong tendency for users to invoke COVID-19 in more toxic posts, a pattern which is aligned with important pandemic events, apparent from their overlap with the BEAST-detected change points.
More specifically, we find three notable patterns. First, the presence of COVID-19 content is strongly predictive of toxicity in climate discussions on both Twitter and Reddit, with the exception of posts that explicitly link to climate science. This pattern is even stronger for Fauci content, which we only examined in general climate Twitter discussions due to their low presence in the other data sets.
Second, increases in the tendency for invoking COVID-19 in toxic climate discussions clearly correspond to times when the link between COVID-19 science and public policy became more salient, such as the lead up to the 2020 U.S. presidential elections and the onset of the Delta variant and its associated policy responses in early summer 2021. 
Third, COVID-19 climate tweets initially started at approximately 0.1 toxicity probability (March--May), which is the same as general climate tweets that do not contain COVID-19 content, but quickly spiked and diverged from the baseline within the first three months (June onward). This temporal pattern is indicative of partisan sorting, whereby new issues enter the public consciousness as nonpartisan ones, but quickly become divisive and politically charged as a result of elite-led polarization \citep{hetherington2001resurgent}, which was strong during this period~\citep{green2020elusive}.

\subsubsection{Probability Models of Climate Post Toxicity}
Next, we looked more closely at content toxicity patterns after the initial incivility escalation period using linear probability models of post-level toxicity.
Specifically, we fitted linear probability models for post-level incivility for the September 2020--August 2021 period, i.e., after initial polarization stabilized as detected by the BEAST models.\footnote{We fitted all models using the \texttt{fixest} package~\citep{berge2018efficient}. Confidence intervals were obtained using the \texttt{marginaleffects} package~\citep{arelbundock2024how}. To account for the bounded nature of probabilities, we show in Supplementary Materials~S8 that our results are robust to using fractional logistic regression models~\citep{papke2008panel}.}

In these models, we controlled for a set of potentially confounding factors. First, we included binary covariates for whether the post mentions international organizations or public figures who are targets of conspiracy theories, and whether the post contains obstructionist or contrarian information about climate change. These are two factors that relate to pre-pandemic sources of harmful engagement with climate content on social media \citep{rojas2024hierarchical,lockwood2018right}.
We also included binary covariates for whether the post is the first post of a conversation, which has a higher chance of being a news or scientific article posted by an institutional account; and whether the posting account was active before the COVID-19 period, because we expect social media platforms to have attracted new users during the pandemic who are more uncivil than existing ones. 
All models include day fixed effects, and we also clustered the standard errors at the appropriate levels. We describe these specifications in Supplementary Materials S6.

As shown in Figure~\ref{fig:toxicity}.B, we find consistent evidence of incivility spillover between climate change and COVID-19 across all three data sets. 
Coefficient estimates from our models of climate posts toxicity show that posts invoking COVID-19 and Fauci have a significantly higher probability of being toxic, although climate science tweets rarely contained Fauci content. Among general climate tweets, these results hold when account fixed effects are included, meaning the observed pattern of COVID-19 posts being more toxic is not simply due to different individuals choosing different things to talk about, but is attributable to an individual-level behavioral tendency to be more toxic when discussing COVID-19. We did not fit account fixed effects models for the scientific tweets data because there is not enough within-user variation in our predictors, nor for the Reddit data because it does not contain user information.

\subsubsection{Survival Models of Toxicity Onset}
Finally, we leverage the sequential structure of conversations to show that general climate tweets that include references to COVID-19 tend to incite greater toxicity in their replies. For this, we fitted Cox proportional hazard models \citep{therneau2000modeling}, where the outcome is the first occurrence of a toxic reply in a Twitter thread from the post-escalation COVID-19 period. 
In our models, we controlled for occurrences of international organizations keywords on the root tweet as a property of the conversation, and occurrences of these keywords, along with Fauci keywords, in subsequent replies were treated as time-dependent covariates. We did not include mentions of Fauci in the root tweet of conversations because this was an extremely rare occurrence. 

Our results are shown in Figure~\ref{fig:toxicity}.C.\footnote{We fitted all models using the \texttt{survival} package~\citep{therneau2024survival-package}. Details of how we implemented the Cox proportional hazards model, including results of test for the proportional hazards assumption~\citep{therneau2000modeling}, are in Supplementary Materials~S7. We show that our results are robust to alternative specifications of what constitutes an English conversation in Supplementary Materials~S9.} First, we find that the number of replies until one exceeds 0.5 toxicity probability is significantly lower in general climate conversation that begin with a tweet mentioning COVID-19. At any given time, replies to these conversations have approximately 14\% higher odds to devolve into incivility compared to conversations that do not begin with COVID-19 content. This pattern however does not hold for climate science conversations. Second, we find that the first toxicity reply in both general climate and climate science conversations are likely to contain mentions of COVID-19 or Fauci, which corroborates our post-level models.

\subsection{COVID-19 Incivility Spillover Activates along Pre-existing Populist Beliefs}\label{sec:international}
We have thus far demonstrated that incivility spilled over between the COVID-19 and climate change domains. We now show evidence that this kind of spillover occurred in part because pre-existing populist anti-internationalist sentiments and conspiracy beliefs provided links between climate policy opposition and COVID-19 vaccine hesitancy.

\begin{table}[!b]
    \centering\footnotesize\renewcommand{\arraystretch}{1.2}
    \begin{tabular}{p{0.48\textwidth} p{0.48\textwidth}} \hline\hline
        Pre-COVID &         Post-COVID \\
        \hline
        Climate change ``scientists'' are not stupid. They are on the biggest ever gravy train. Politicians are the stupid ones for allowing them and UN bodies to rip us off. All part of the Soros funded NWO [New World Order].
        &
        We won't be in deep shit because of non-existent climate change. We'll be in deep shit because dumb fucks like these people are in charge, and are pushing a hoax for more total control. Klaus Schwab%
        , Fauci, Gates, etc.
        \\ \hline

    \end{tabular}
            \caption{\textbf{Examples of uncivil climate tweets that mention international organizations before and after the onset of COVID-19 pandemic.} Toxicity probabilities of these tweets exceed 0.75. Tweet texts have been edited for typos and formatting.}\label{tab:org-tweets}
\end{table}

Since the 21\textsuperscript{st} century populist surge~\citep{mudde2019far}, there has been decreasing support for international organizations and globalist policies, a sentiment manifesting most strongly among groups adversely affected by a globalized economy ~\citep{bearce2019popular}. Specific to climate policy, with the growing political salience of the climate issue, far-right populist parties have started taking an anti-internationalist opposition toward international climate governance~\citep{lockwood2018right,schworer2024climate,witajewska2024politicization}. At the extreme, these sentiments devolve into conspiracy theories about international elites trying to exert global control~\citep{castanho2017elite} -- such as the example in Table~\ref{tab:org-tweets} about George Soros and the United Nations.

Similar patterns exist in the public health domain, where conspiracy beliefs, including those about international elites, tend to align with anti-vaccination attitudes~\citep{hornsey2018psychological}. These were fringe positions, but became popular during the pandemic. On social media, anti-vaccination and anti-internationalist conspiracy theories became rampant~\citep{darius2021disinformed, erokhin2022covid}. Offline, anti-vaccine individuals in the U.K. often held pseudo-scientific, populist, and conspiracy beliefs~\citep{holford2023psychological}, and a sizable segment of the U.S. population viewed the World Health Organization negatively during the pandemic~\citep{bayram2021trusts}.

Did the COVID-19 pandemic elevate existing populist sentiments against climate science and policies? Based on prior research showing that cognitively coherent beliefs reinforce one another~\citep{taber2006motivated} -- which has been demonstrated in belief systems of conspiracy theories \citep{enders2021conspiracy} -- we expect overlapping antagonistic populist beliefs toward international elites and scientists in the realms of public health and climate change to have served as a pathway for spillover between the two global crises, leading to greater antagonism toward international organizations in climate discussions during the pandemic.

To explore this possibility, we examined the temporal dynamics, before and during the pandemic, of toxicity differences between climate tweets that referenced international organizations and those that do not.
Specifically, we fitted a model for tweet toxicity using our general climate Twitter data set for our entire observation period (February 1, 2019--August 26, 2021) so we can examine second differences based on time period~\citep{mize2019best}. 
In the model, we interacted the international organization keyword term with every quarter in the data (Q1 2019--Q3 2021), then compared the coefficient of the international organization term from each quarter to the pre-pandemic baseline (Q4 2019). 
To ensure the increased incivility during the pandemic is not directly due to COVID-19 tweets, we removed them from this analysis, but our results are robust to whether these tweets are included, which we show in Supplementary Materials S10.

Because anti-internationalist sentiments lead to greater incivility in climate discussions, if the anti-internationalist sentiments activated by COVID-19 spilled into the climate domain, there should be greater toxicity in climate change discussions about international organizations compared to the baseline, even in tweets not directly concerning the pandemic. As the onset of the pandemic was exogenous to climate change in the short term, observed temporal differences can be attributed to the COVID-19 shock.

\begin{figure}[!t]
    \centering
    \includegraphics[width=0.5\textwidth]{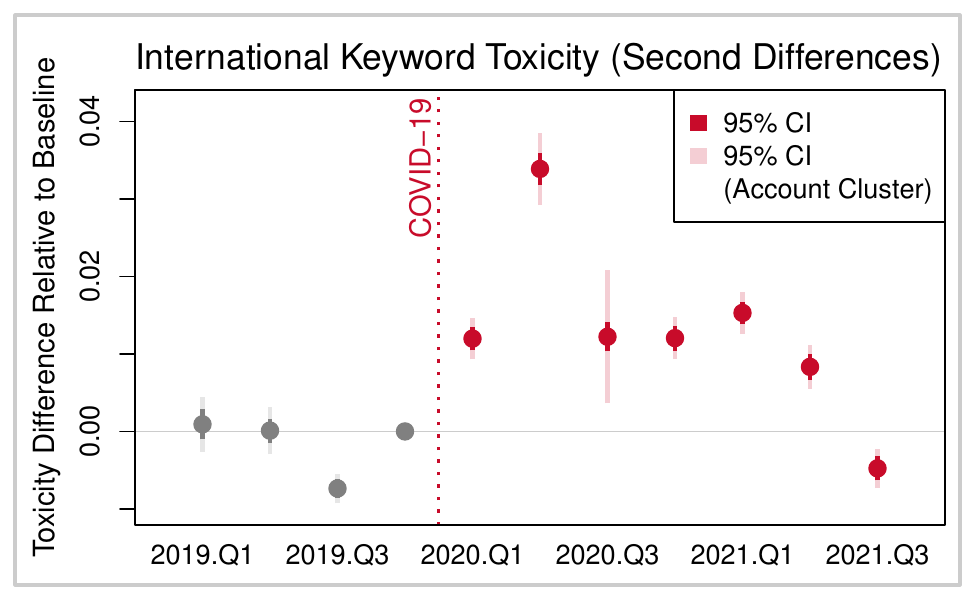}
    \caption{\textbf{Quarterly difference in toxicity between tweets containing references to international organizations and those that do not, compared to the baseline quarter (2019 Q4).} The lighter confidence intervals are computed from standard errors clustered at the Twitter account level. 
    }
    \label{fig:io-toxicity}
\end{figure}

As we show in Figure~\ref{fig:io-toxicity}, we find a significant increase in toxicity toward international organizations in climate discussions during the COVID-19 period starting Q1 2020, when compared to the baseline period of Q4 2019. The difference in toxicity probability before the pandemic, on the other hand, was relatively stable. The highest spike occurred in Q2 2020, immediately following the World Health Organization declaring COVID-19 to be a global pandemic, and not until Q3 2021 did the difference fall to pre-pandemic levels. These findings provide clear evidence that the nature of climate discussions about international organizations changed with the pandemic.

\section{Conclusion}
In this study, we showed that affective polarization surrounding COVID-19 spilled over to how the public engages with climate change and climate science on social media. References to COVID-19 were strongly associated with incivility in climate discussions. On the whole, this relationship was responsive to landmark pandemic era events, as incivility increased when the role of science in policymaking was made more salient by COVID-19 policies. 
We further showed that populist sentiments against international organizations and elites provided a link between climate policy opposition and COVID-19 vaccine hesitancy. This provides evidence that antagonism spillover occurred in part through existing political cleavages, which supports our understanding that this spillover is part of a larger political sorting process that produced entrenched belief systems governing belief formation and political communication. 

%

Our findings raise important questions for future research.
In our context of public health and climate change, linkages via pre-existing cleavages led to decidedly harmful outcomes. Given our two issues, we are unable to speak to what would happen in cases where the pre-existing synergistic pathway is weaker or nonexistent. Given that not all ideas are as strongly linked by political attitude systems or conspiracy belief systems, are there limits to incivility spillover based on how much conceptual distance there is between issues? Related to this, it would be valuable both theoretically and practically to jointly assess our findings with those from \citet{bergquist2023politics}, which instead show the possibility for COVID-19 and climate change to yield synergistic policy support. What are the conditions under which different synergies become activated, or is it a matter of different segments of the population reacting to different cues?

While the COVID-19 pandemic and its public salience has subsided, legacies of pandemic era politics continue today. Science being contested as part of the political process, whether it be climate change or public health, was apparent with the political vitriol surrounding Anthony Fauci. It continues with politically-motivated appointments to science-related posts following government transitions~\citep{tanne2024trump}. Trust in scientists remains relatively low at levels similar to the pandemic period~\citep{pew2024public}, and recent research shows worrying signs of behavioral spillover between COVID-19 and post-pandemic vaccination attitudes~\citep{lunz2024covid}. With climate change remaining ever as important, it is imperative to better understand how spillover among concurrent crises might lead to system-wide entrenchment of affectively polarized attitudes.

Incivility has adverse implications for science communication, manifesting as a vicious cycle between decreasing trust and worsening intergroup conflict.
%
By demonstrating the existence of affective polarization spillover and how it activated along existing political cleavages, our study adds to the literature on how COVID-19 impacted the public's engagement with the climate issue, which has thus far focused on how concurrent crises can lead to shifts in their relative issue saliency~\citep{smirnov2022covid,sisco2023examining} and the potential for policymaking synergies between them~\citep{bergquist2023politics}. 
Beyond public health and climate change, our findings demonstrate that conceptual synergies rooted in populist sentiments can lead to harmful antagonistic behaviors spilling over across different public policy domains that are strongly informed by science, 
reiterating the dangers of entrenched group cleavages on how the public engages with issues at the science-policy nexus.


\singlespacing\normalsize
\section*{CRediT Author Statement} 
\textbf{Hasti Narimanzadeh:} Conceptualization, Methodology, Software, Validation, Formal Analysis, Investigation, Data Curation, Writing - Original Draft. 
\textbf{Arash Badie-Modiri:} Conceptualization, Methodology, Software, Validation, Formal Analysis, Investigation, Data Curation, Writing - Original Draft, Visualization. 
\textbf{Iuliia Smirnova:} Conceptualization, Validation, Investigation, Data Curation. 
\textbf{Ted Hsuan Yun Chen:} Conceptualization, Methodology, Validation, Formal Analysis, Investigation, Writing - Original Draft, Visualization, Supervision, Funding Acquisition.

\bibliography{climate}

\end{document}


\singlespacing

\maketitle\thispagestyle{empty}

\tableofcontents


\setcounter{figure}{0}
\setcounter{table}{0}
\setcounter{equation}{0}

\clearpage
\appendix
\normalsize
\onehalfspacing
\newcounter{sisection}
\setcounter{sisection}{1}

\renewcommand{\thesection}{S{\arabic{sisection}}}
\renewcommand{\thefigure}{\thesection.\arabic{figure}}
\renewcommand{\thetable}{\thesection.\arabic{table}}
\renewcommand{\theequation}{\thesection.\arabic{equation}}

\counterwithin{figure}{section}
\counterwithin{table}{section}

\section{Twitter Data Collection}\label{app:twitter-collect}
We used Twitter's Academic API to collect our Twitter data. To collect the general climate tweets, we queried the API for tweets with hashtags and keywords from Table~\ref{tab:collection} for the February 1, 2019--August 26, 2021 period (missed data from February 4, 2021). We excluded tweets containing ``political climate'' and ``Percip:''. The latter is frequently used by bots that automatically tweet regional weather forecast. This yielded 38.4 million English tweets.

To identify all tweets that reference (through hyperlinks) climate science publications, we first used Altmetric's Explorer API to filter scientific publications by whether their titles contained one of our keywords (Table~\ref{tab:collection}), then used Altmetric's Details Page API to obtain IDs of all tweets that referenced these publications. We used Twitter's Academic API to collect these tweets. This yielded 259,000 English tweets.

\begin{table}[!htb]
    \centering\renewcommand{\arraystretch}{1.2}\footnotesize
    \begin{tabular}{p{0.10\textwidth} p{0.87\textwidth}} \hline\hline
        Type & String  \\ \hline

        Hashtags & climatehoax, globalwarming, climateneutrality, climatecrisis, climatebrawl, climaterisk, chooseforward, climateemergency, climatestrike, climatechange, climatefriday, climatescience, actonclimate, climatehysteria, climatestrikeonline, climatetwitter, climatejustice, climate, fridaysforfuture, fridays4future, schoolstrike4climate, facetheclimateemergency, climatetwitter, climatetech, globalwarminghoax, gretathunberg, parisagreement, nomoreemptypromises \\
        Keywords & climate, global warming, greenhouse gas, greenhouse emission, paris agreement \\
        \hline
    \end{tabular}
        \caption{\textbf{Hashtags and keywords used for filtering climate change tweets.}}\label{tab:collection}
\end{table}

We then used Twitter's API to collect the conversation trees stemming from our climate and climate science tweets. Specifically, for all root tweets that have at least one reply from a different user, we obtained all replies to the root tweet and all replies to replies, infinitely deep. For the general climate change data, we collected a stratified random sample of approximately 2100 conversations per week, and the entire set of climate science conversations.

All Twitter data sets were subset to only tweets tagged as English by Twitter and flagged as containing English by Perspective's multilingual detector.
Table~\ref{tab:dataset-users} shows the distribution of the number of users that were active before and after the COVID-19 onset.

\begin{table}[htb!]
    \centering\footnotesize\renewcommand{\arraystretch}{1.2}
    \begin{tabular}{r | c c}\hline\hline
        User Subset & Climate General & Climate Science\\ 
        \hline
        Only before the pandemic & 2.0M & 28K \\
        Only during the pandemic & 2.1M & 37K \\
        During the entire period & 1.3M & 14K \\ \hline
        Total users & 5.5M & 79K \\ \hline
    \end{tabular}
\caption{\textbf{User counts by their activity before and during the pandemic.}}\label{tab:dataset-users}
\end{table}

\clearpage
\stepcounter{sisection}
\section{Reddit Data Collection}\label{app:reddit-collect}
For our Reddit analysis, we used the Reddit Climate Change Dataset provided by \citet{socialgrep2022reddit}. This data set consists of approximately 4.6 million Reddit comments posted before September 2022 that contain ``climate change''. We removed comments posted outside our time window and any comments where Perspective API did not listed English among the languages present. We also removed comments containing phrases indicating that they were written by bots. Specifically, we filtered the comments on a set of common phrases used by bots on Reddit to perform automatic tasks, e.g.,~``I am a bot'' or ``This action was performed automatically'' (Table~\ref{tab:reddit-bot-phrases}), as well as excluding all comments from the subredditsummarybot because it mostly consist of automatic submissions. Our final Reddit data set contains 2.1 million comments.

\begin{table}[!htb]
    \centering\footnotesize\renewcommand{\arraystretch}{1.2}
    \begin{tabular}{l}\hline\hline
         Filter phrase  \\
         \hline
         I am a bot \\
         I'm a bot \\
         This action was performed automatically \\
         This message was posted by a bot. \\
         This book has been suggested \\
         This comment was left automatically (by a bot). \\
         You can summon this bot any time in \\
         I detect haikus. And sometimes, successfully. \\ \hline
    \end{tabular}
        \caption{\textbf{Phrases used for filtering automatic Reddit comments.}}\label{tab:reddit-bot-phrases}
\end{table}

\clearpage
\stepcounter{sisection}
\section{Topic Classification}\label{app:topic-keywords}

Based on the presence or absence of specific keywords, we labeled all posts as containing content from the following categories: 1) COVID-19, 2) Anthony Fauci, and 3) international organizations. Table~\ref{tab:keyword-categories} contains our keywords.

To lower the probability of false positives, Twitter handles and the keyword ``gates'' were only considered if surrounded by word boundaries. Keywords shown in all caps were searched case sensitively and surrounded with word boundaries. The keyword ``WHO'' was specifically searched case sensitively and surrounded with word boundaries, and only matched if 
at least 70\% of the tweet was uppercase. 
This condition was set to avoid tweets simply containing the interrogative ``who'' written in all caps.
All other keywords were searched case insensitively while disregarding word boundaries, allowing us to capture them in hashtags or with affixes.

\begin{table}[!htb]
    \centering\footnotesize\renewcommand{\arraystretch}{1.2}
    \begin{tabularx}{\textwidth}{l X} \hline \hline
Category & Keywords \\ \hline
COVID-19 & coronavirus, corona virus, covid,
        covid19, covid-19, mask, pandemic,
        lockdown, wuhan, sars-cov-2, sarscov2,
        flatten the curve, flattening the curve, flatteningthecurve,
        flattenthecurve, hand sanitizer, handsanitizer,
        social distancing, socialdistancing, work from home,
        workfromhome, working from home, workingfromhome,
        mrna, vax, vaccin, pfizer,
        the jab, jabbed, jabbing,
        moderna, astrazeneca, biontech, sinovac,
        sinopharm, johnson \& johnson, johnson\&johnson\\

Anthony Fauci & fauci\\
International Organizations & WHO, @who,
        w.h.o, world health org, WEF, @wef,
        w.e.f, world economic forum, Davos,
        UN, @un, u.n, united nations,
        gates, billgates, @billgates, @gatesfoundation,
        soros, @georgesoros\\ 
\hline
    \end{tabularx}
        \caption{\textbf{Keywords used for determining mentions of different topics in tweet texts.}}\label{tab:keyword-categories}
\end{table}

\subsection{Climate Obstructionist Content}
To determine whether tweets contain obstructionist claims about climate change, we used the Augmented CARDS (Computer Assisted Recognition of Denial \& Skepticism) model presented in~\citep{rojas2024hierarchical}. The model contains a binary classifier layer based on a DeBERTa neural language model \citep{he2020deberta}, designed to detect tweets as contrarian or otherwise, specifically, if it contains any of the five main types of contrarian claims: 1) global warming is not happening, 2) humans are not causing global warming, 3) climate change is not bad, 4) climate solutions will not work, or 5) climate movement or science are unreliable.

\clearpage
\stepcounter{sisection}
\section{BEAST Time Series Details}\label{app:beast}
BEAST is a time series decomposition algorithm designed to analyze nonlinear temporal dynamics across multiple timescales while taking into account seasonal behavior (e.g.,~weekly trends) without manually specifying the model. It does so by simultaneously estimating the temporal trends and cyclical patterns that make up an observed time series, and the change points in both of these dynamic processes. 
Specifically, it decomposes the observed time series into a trend that is linear between change points (i.e., a series of piecewise linear trends) and a cyclical pattern that follows a harmonic function whose parameters are constant between change points. 
Instead of opting for a single ``best'' model, BEAST uses Bayesian model averaging of multiple competing models to obtain the point and uncertainty estimates of model parameters -- the trend, the harmonic function's parameters, and the number and location of change points.

In our analysis, our time series are at the day level. We specified the cyclical pattern to be every seven days, given our expectations about Twitter activity patterns (e.g., there to be less activity on the weekends). When estimating change points, we specified there to be a maximum of fifteen change points (which was never reached), and that any two change points must be at least four weeks apart.


\clearpage
\stepcounter{sisection}
\section{Sample Sizes for All Fitted Models}\label{app:lpm}
\autoref{tab:samplesize} contains the sample sizes for all models presented in the main text.

\begin{table}[!ht]
    \centering\footnotesize\renewcommand{\arraystretch}{1.2}
\begin{tabular}{l r} \hline\hline
    Model    & Sample Size \\ \hline
     \multicolumn{2}{l}{\textbf{Linear Probability Model}} \\
     Climate General & 14,640,829 \\ 
     Climate General (User Fixed Effects) & 14,640,829 \\
     Climate Science & 109,447 \\
     Reddit & 756,773 \\
     International Organizations & 37,010,541 \\ \hline

     \multicolumn{2}{l}{\textbf{Cox Proportional Hazards Model*}} \\ 
     Climate General & 233,208  \\
     Climate Science & 39,384 \\
     \hline
    \end{tabular}
            \caption{\textbf{Sample sizes for all models.} 
            The reported sample size for the Cox proportional hazards models are the average observations across all bootstrap samples.
            }\label{tab:samplesize}
\end{table}

\clearpage
\stepcounter{sisection}
\section{Linear Probability Models Details}\label{app:lpm}
We fitted separate linear probability models of climate post toxicity for all three of our data sets -- general climate tweets, climate science tweets, and climate Reddit posts. We focused on the COVID-19 period after the initial incivility escalation (August 14, 2020--August 26, 2021) as detected by the BEAST models. In all three models, we included day fixed effects. In all models, we clustered the standard errors at the day level, and also at the subreddit level for the Reddit analysis. We additionally included account fixed effects and account clustering for the general climate tweets data set in a separate model. Fitting all models using fractional logistic regression~\citep{papke2008panel} yielded substantively similar results, which we show in Section~\ref{app:fraclogit}.


\clearpage
\stepcounter{sisection}
\section{Cox Proportional Hazards Model Details}\label{app:coxph}
We modelled the time to onset of toxicity in Twitter climate conversations using the Cox proportional hazards model, focusing on whether this varied by the COVID-19 variables we are interested in.

\subsection{Model Specification}
We analyzed the general climate conversations and climate science conversations separately. For both sets of analysis, we used all English conversations August 14, 2020 onward from the respective data sets. 

In the models, we included a mixture of both thread-level covariates, i.e., occurrences of certain keywords in the root tweet, and reply depth-level covariates, i.e., occurrences of these keywords in reply tweets. Specifically, we included covariates for whether the root tweet contains COVID-19 keywords and international organization keywords, and covariates for whether specific replies contain COVID-19, Fauci, and international organization keywords.

Finally, to prevent violations of the proportional hazards assumption from these models, we stratified our models by whether the conversation's root tweet was already toxic.

\subsection{Data Construction}
In its raw form, a conversation is composed of replies to the root tweet, and replies to these replies, up to an infinite depth. At any point when there are multiple replies to the same tweet, the reply tree branches into different threads. To prepare this conversation data for the Cox proportion hazards model, we take the following steps.

\begin{itemize}
    \item Each thread, or branch of the conversation tree defined by its unique terminal tweet, were treated as separate entities that we observe over time.
    \item Threads started by the automated account @wikipediachain, which document random walks through Wikipedia links, were removed.
    \item Tweets with toxicity probability that are not able to be estimated by Perspective API and all subsequent replies in its thread are removed, because we are unable to assess how they affect subsequent replies. (The Perspective API only supports the following languages: Arabic, Chinese, Czech, Dutch, English, French, German, Hindi (including when written with Latin alphabet), Indonesian, Italian, Japanese, Korean, Polish, Portuguese, Russian, Spanish, and Swedish; and is unable to parse tweets containing only non-textual material, e.g., images, videos, or links.) The Perspective API returned NA values for 3.4\% of general climate tweets and 6.5\% of climate science tweets.
    \item Consecutive tweets from the same author, which generally appear because of Twitter character limits, were combined into a single tweet.
    \item Each tweet was classified as toxic if its toxicity probability was at least 0.5.
    \item For each thread, tweets after the first occurrence of a toxic tweet were dropped, because we are modelling toxicity onset. 
    \item If no toxic tweet was found in the entire chain, the thread was treated as right-censored. 
    \item If a tweet was deleted or made private somewhere along the chain, the thread was likewise treated as right-censored. In terms of systematic deletions that could impact our data, while it is difficult to know the impact of Twitter's moderation policies, it was reported that as of March 2021 (i.e., approximately seven months before our data collection started), Twitter had deleted only approximately 8400 tweets for COVID-19 misinformation \citep{twitter2021covid}.
    \item All root tweets were removed, because we are modelling reply toxicity.
\end{itemize}

These steps yielded a data set with thread-reply depth observations. Each observation has thread-level covariates (e.g., COVID-19 content in root tweet) and reply depth-level, i.e., ``time-varying'', covariates (e.g., COVID-19 content in reply tweet).

\subsection{Estimation}
As threads from the same conversation are not independent from one another due to sharing tweets higher up in the conversation (i.e., closer to the root tweet), we based our analysis on a resampling method. In each realization, we randomly sampled one thread from each conversation. This means that no two threads in the same realization come from the same conversation, and therefore do not shared any tweets. This resampling eliminates the dependence between chains from the same conversation, and also ensures that the proportional hazard assumption is not violated due to presence of extremely large conversations.

Using the \texttt{survival} package \citep{therneau2024survival-package} in \texttt{R} \citep{r2024}, we fitted 20,000 resampled realizations of our data. In our results, we report the mean coefficient estimates as the point estimate and the 2.5\textsuperscript{th} and 97.5\textsuperscript{th} percentile values as the confidence interval.

\subsection{Proportional Hazards Assumption Diagnostics}
We assessed whether our models satisfy the proportional hazards assumption using the Schoenfeld test \citep{therneau2000modeling}. We conducted the test for all fitted models and report the proportion of realizations that pass the test in Table~\ref{tab:schoenfeld}.

\begin{table}[ht!]
    \centering\footnotesize\renewcommand{\arraystretch}{1.2}
    \begin{tabular}{r | c c} \hline\hline
    & \multicolumn{2}{c}{Proportion of Bootstraps Not Failing Schoenfeld Test} \\ 
            & All Tweets & Root Scientific Tweets  \\
        \hline
        COVID-19 root           & 0.99 & 0.89 \\
        COVID-19 keywords       & 0.76 & 1.00 \\
        Fauci keywords          & 1.00 & 1.00 \\
        International root      & 0.98 & 0.85 \\
        International keywords  & 1.00 & 1.00 \\
        \hline
    \end{tabular}
        \caption{\textbf{Cox proportional hazards assumption assessment.} Proportion of bootstraps not failing the Schoenfeld test with $p > 0.05$.}\label{tab:schoenfeld}
\end{table}


\clearpage
\stepcounter{sisection}
\section{Fractional Logistic Regression}\label{app:fraclogit}
\begin{figure}[!h]\centering
    \includegraphics[width=0.5\linewidth]{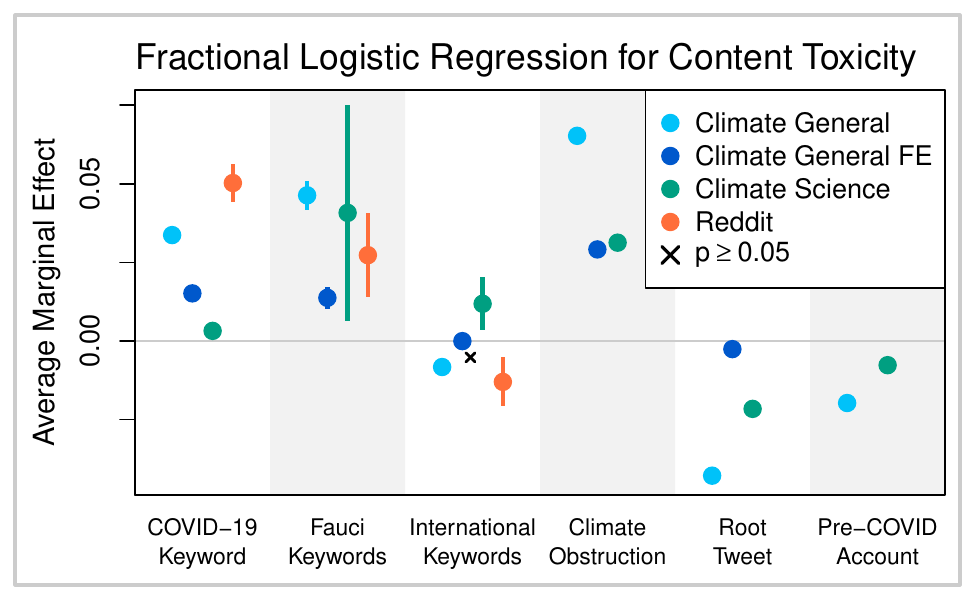}
        \caption{\textbf{Robustness check of main results using fractional logistic regression models.} Average marginal effect estimates for toxicity probability, corresponding to results from Figure~1.B from the main text.}\label{fig:robust-fraclogit}
\end{figure}

Linear probability models often provide an adequate approximation of non-linear models, especially when the range of outcome probabilities are not extreme. Still, we fit additional fractional logistic regression models \citep{papke2008panel} with the same model specifications as what we used for our main analysis to show, in Figure~\ref{fig:robust-fraclogit}, that our results are robust to accounting for the bounded nature of probabilities.

\clearpage
\stepcounter{sisection}
\section{Cox Proportion Hazards Model without Subsetting to English Replies}\label{app:coxph-robust}

Our standard approach to data preprocessing is to subset to only English posts. For our Twitter conversation analyses using the Cox proportional hazards model, this means removing the first appearance of a non-English tweet in a thread then truncating the rest of the thread, as we cannot assess how the non-English replies affected subsequent replies. We show that our findings regarding COVID-19 content in conversations are robust to an alternative specification, where we subset to conversations beginning with an English root tweet, do not limit replies to only English ones. 

\begin{figure}[!h]
    \centering
    \includegraphics[width = 0.5\textwidth]{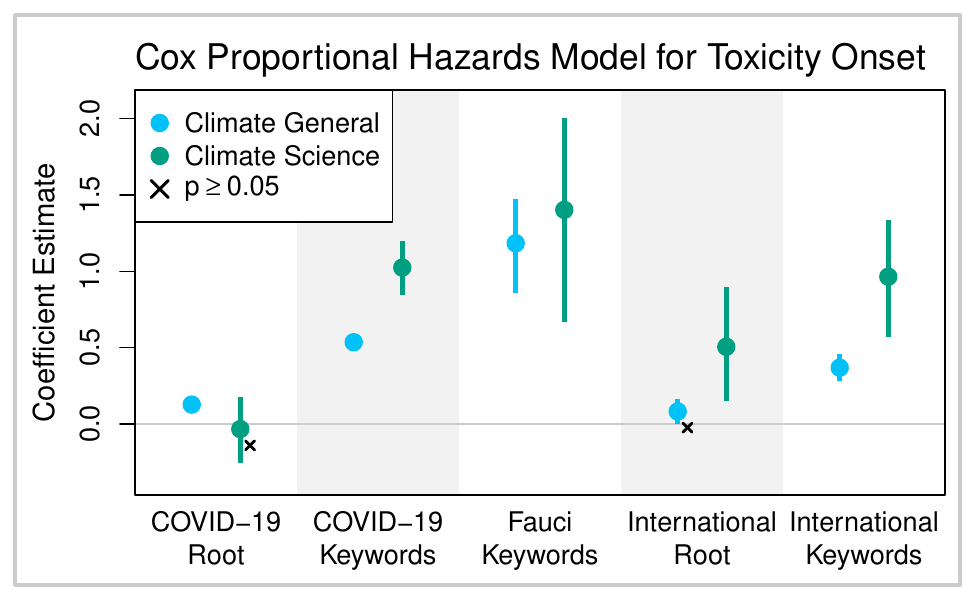}
        \caption{\textbf{Robustness check of main Cox proportional hazards model results using full replies instead of only English ones.} Figure corresponds to Figure~1.C from the main text.}\label{fig:coxph-robust}
\end{figure}

\clearpage
\stepcounter{sisection}
\section{COVID-19 Tweets in the Anti-internationalist Analysis}\label{app:io-robust}
\begin{figure}[!h]
    \centering
    \includegraphics[width = 0.5\textwidth]{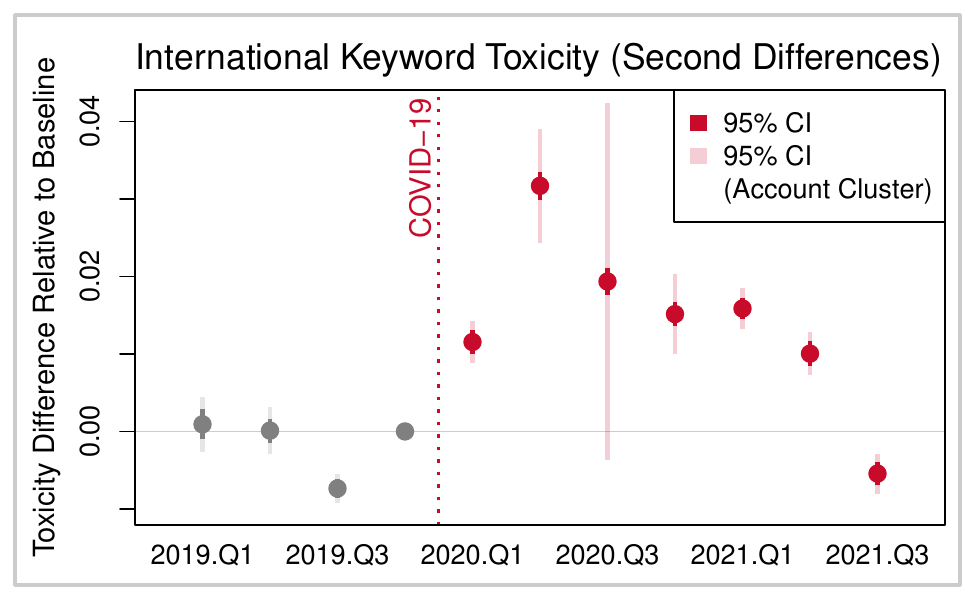}
        \caption{\textbf{Robustness check of toxicity probability of international organization tweets over time when including tweets with overlapping COVID-19 content.} The lighter confidence intervals are computed from standard errors clustered at the Twitter account level. Figure corresponds to Figure~2 from the main text.}\label{fig:io-toxicity-covid}
\end{figure}

When assessing the impact of COVID-19 on the relationship between anti-internationalist sentiment and toxicity in climate discussions, we examined only tweets that do not directly engage with COVID-19 to ensure our results are not driven purely by the toxicity of COVID-19 content. Still, we show, in Figure~\ref{fig:io-toxicity-covid}, that our results are robust to the inclusion of COVID-19 tweets.

\clearpage
\stepcounter{sisection}
\bibliography{climate}